\documentclass[preprint,aps,prb,showpacs,secnumarabic,nobibnotes,floatfix,preprintnumbers,amsmath,amssymb,superscriptaddress]{revtex4-1}

\usepackage{graphicx}
\usepackage{dcolumn}
\usepackage{bm}
\usepackage{bibtopic}
\usepackage{multirow}
\usepackage{subfigure}
\usepackage{xcolor}
\usepackage[normalem]{ulem}

\begin{document}

\title{Effect of co-existing crystal structures on magnetic behavior of Ni$_{2}$Mn$_{1+x}$Sn$_{1-x}$ magnetic shape memory alloy }
  
            \author{Soumyadipta Pal}
      
      \affiliation{Department of Condensed Matter Physics and Material Sciences, S N Bose National Centre for Basic Sciences, Block-JD, Sector-III, Salt Lake, Kolkata - 700106, West Bengal, India.}
      \affiliation{Department of Physics, Calcutta Institute of Technology, Banitabla, Uluberia, Howrah - 711316, West Bengal, India.}

      \author{Sandeep Singh}
      
      \affiliation{Department of Condensed Matter Physics and Material Sciences, S N Bose National Centre for Basic Sciences, Block-JD, Sector-III, Salt Lake, Kolkata - 700106, West Bengal, India.}
      
      \author{Chhayabrita Maji} 
      
      \email{Email: chhayabrita@gmail.com} 
      \thanks{Corresponding author; Previous publication name: C.Biswas}
      
      \affiliation{Department of Condensed Matter Physics and Material Sciences, S N Bose National Centre for Basic Sciences, Block-JD, Sector-III, Salt Lake, Kolkata - 700106, West Bengal, India.}
      \affiliation{Department of Materials Science, Indian Association for the Cultivation of Science, 2A \& 2B Raja S.C. Mullick Road, Jadavpur, Kolkata - 700032, West Bengal, India.}

\begin{abstract}  

The temperature dependent crystal structure analysis in martensitic phase of Ni$_{2}$Mn$_{1+x}$Sn$_{1-x}$ ($x$ = 0.40 and 0.44) magnetic shape memory alloy is performed using X-ray diffraction. The martensitic phase consists of 4-Layered and 14-Layered orthorhombic structure. The phase fraction of 4-layered and 14-layered orthorhombic structure change with temperature. Interestingly, the magnetic property also changes with temperature and corresponds to structural change temperature. Thus, the exchange coupling constant between original Mn and substituted Mn is calculated, which shows that the 4-layered and 14-layered orthorhombic structures have different strength of antiferromagnetic coupling. The analysis explain the origin of spin glass behavior and unusual exchange bias effect in these systems under zero-field-cooling.
  
\end{abstract}

\pacs{81.30.Kf,83.10.Tv,75.50.Cc,81.05.Bx}

\maketitle

\section{Introduction}

The new class of off-stoichiometric magnetic shape memory alloys (MSMA) that exhibit shift of martensitic transition to lower temperature with applied magnetic field has gained adequate scientific attention due to shift induced attractive properties like giant magnetocaloric effects \cite{Krenke_Nature_2005}, giant magneto-resistance \cite{Sandeep_APL_2011}, magnetic shape memory effects \cite{Imano_Nature_2006} and exchange bias effects \cite{Wang_PRL_2011}. These fascinating properties make the alloys a potential candidate for applications in solid-state magnetic refrigeration \cite{Krenke_Nature_2005}, magnetic actuators \cite{Karaca_AdvFunctMater_2009,Jiles_Sensor_2003} and magnetic sensors\cite{Jiles_Sensor_2003} etc. This class of material are very important for replacing present way of cooling with hazardous gases. These MSMA exhibit giant magnetocaloric effect, also, with applied pressure that are comparable to reported present material for solid-state refrigeration \cite{Nature_Manosa2010}. Moreover, these alloys offer cost effective technology.

Since the pioneering work on exchange bias (EB) by Meikleohn and Bean in 1956 \cite{Meiklejohn_Bean_EB_1956}, this phenomenon has been explored extensively due to its various applications in spintronics, magnetic recording and sensors devices \cite{Jiles_Sensor_2003,Kilian_JAP_2000,Mori_TSF_2012,Bainsla_App-phy-rev_2016,Nikolaev_JAP_2008}. It was usually observed in materials with interface between different magnetic phases such as ferromagnetic (FM)-antiferromagnetic (AFM), FM-spin glass (SG),
AFM-ferrimagnetic (FI), and FM-FI phases \cite{Meiklejohn_Bean_EB_1956,Ali_Nature_2007,Giri_JPCM_2011,Razzaq_JAP_1984,Cain_JAP_1990} during cooling with magnetic field to the low temperatures. The conventional EB effect is ascribed to FM unidirectional anisotropy formed at the interface between different magnetic phases in the process of field cooling (FC) from higher temperature. Interestingly, an unusual EB effect under zero-field-cooling (ZFC) has been obtained in many off-stoichiometric Ni-Mn-Z (Z = Sb, Sn and In) Heusler alloys \cite{Wang_PRL_2011,Li_APL_2007,Khan_JAP_2007,Xuan_APL_2010,Wang_JAP_2008,Khan_APL_2007,Wang_JAP_2012}. The Wang \textit{et. al.} \cite{Wang_PRL_2011,Wang_JAP_2012} observed a large EB effect in Ni-Mn-Z (Z = In and Sn) alloys after zero-field cooling (ZFC) from an unmagnetized state. The occurrence of EB effect in ZFC state is attributed to the superferromagnetic (SFM) unidirectional anisotropy below the blocking temperature (T$_{B}$) at low temperature. This is similar to FM-AFM interface with unidirectional FM spins formed after FC in the conventional EB effect. According to Wang \textit{et. al.} \cite{Wang_JAP_2012} probable origin is purely magnetic and does not originate from structural modifications in martensitic phase. On the other hand, Sharma \textit{et. al.} \cite{Sharma_APL_2015} related the Mn-Mn distance with the complex magnetic state in martensitic phase at low temperatures due to co-existence of FM and AFM exchange interactions that lead to magnetic frustration. These conflicting opinions require a detailed insight into the crystal structure of Ni-Mn-Sn Heusler alloys and careful analysis to check the relation between crystal structure and magnetic behavior. In Ni-Mn-Sn martensitic phase FM and AFM magnetic phases co-exist. The AFM coupling occurs between Mn at original site (Mn1) and Mn at Sn site (Mn2) \cite{Pal_AIPProced_2014,Ye_PRL_2010,Kanomata_MaterSciForum_2008,Brown_JPCM_2006}. This occurs due to the decrease in Mn-Mn bond length. Thus, crystal structure change in martensitic phase might play an important role in producing unusual exchange bias effect under ZFC.        

In this manuscript, the crystal structure as a function of temperature for Ni$_{2}$Mn$_{1+x}$Sn$_{1-x}$ ($x$ = 0.40 and 0.44) alloys are investigated till 10 K under ZFC condition. In the martensitic phase the co-existence of 4-layered (4L) and 14-layered (14L) orthorhombic crystal structure is revealed. Interestingly, the phase fraction of 4L and 14L crystal structures changes with temperature. The changes occur at temperatures where magnetic transitions also occur. This implies that there is a possible relation between crystal structure and magnetic behavior. In order to establish the relation, the magnetic ground state of 4L and 14L structures, via the magnetic exchange parameter (J) calculation, between first nearest neighbour Mn1-Mn2 are performed. The calculation reveals the AFM exchange interaction with different strength of AFM coupling for 4L and 14L orthorhombic crystal structure. Important to note that not only FM-AFM domains co-exists, but also different strength of AFM coupling exists. Thus, frustrated state of martensitic phase arises that causes spin-glass-like behavior of Ni$_{2}$Mn$_{1+x}$Sn$_{1-x}$ ($x$ = 0.40 and 0.44) and leads to exchange bias phenomena even under ZFC. Thus, crystal structure of martensitic phase effects the magnetic behavior of Ni-Mn-Sn magnetic shape memory alloys. 

\section{Experimental and Theoretical Method}

\begin{figure}[htb]
\centering
\includegraphics[angle=0,width=10cm,height=6cm]{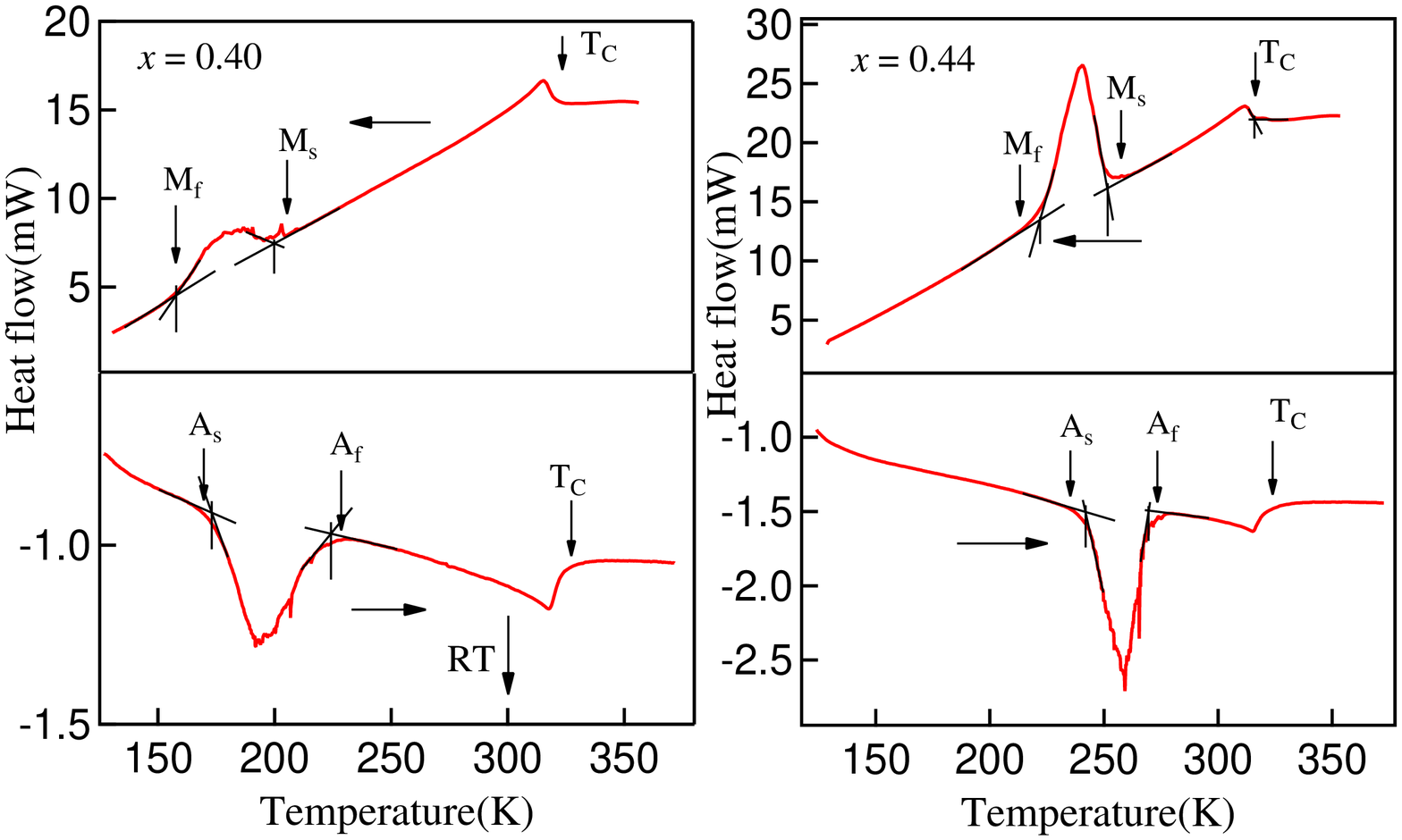}
\caption{Exothermic and endothermic behavior of Ni$_{2}$Mn$_{1+x}$Sn$_{1-x}$ (0.40 and 0.44) using differential scanning caloriemetry during cooling and heating, respectively.}
\label{fig1}
\end{figure}

The polycrystalline ingots of Ni$_{2}$Mn$_{1+x}$Sn$_{1-x}$ ($x$ = 0.40 and 0.44) alloys are prepared by arc melting appropriate amount of high purity ($\geq$ 99.99 \%) constituent elements under argon atmosphere and were annealed at 1173 K (24 h) with subsequent quenching to ice water. The alloys are characterized as mentioned in Ref. 2. The structural and magnetic transition temperatures are determined by differential scanning calorimetry (DSC) measurements (fig. \ref{fig1}).
The values of the transition temperatures are tabulated in Table \ref{Table1}. 
\begin{table}[htbp]
\centering
\caption{The structural and magnetic transition temperatures of $x$ = 0.40 and 0.44, obtained from DSC and magnetization. The maximum error is $\pm$10 K. \cite{Sandeep_APL_2011}}
\footnotesize
\resizebox{8cm}{!}{
\begin{tabular}{cccccccc}
\hline
\hline
Composition& M$_{s}$ & M$_{f}$ & A$_{s}$ & A$_{f}$ & T$_{c}^{M}$ & T$_{c}^{A}$ & T$_{B}$\\ 
($x$)&  (K) & (K) & (K) & (K) & (K) & (K) & (K)\\
\hline
0.40& 189 & 159 & 173 & 208 & - & 326 & 70\\
\hline
0.44& 260 & 229 & 245 & 275 & - & 323 & 80\\
\hline
\hline
\end{tabular}}
\label{Table1}
\end{table}
Below room temperature, $x$ = 0.40 and 0.44 transform from ferromagnetic austenitic phase (AP) to mixed magnetic martensitic phase (MP) where co-existence of FM and AFM coupling is reported \cite{Pal_AIPProced_2014,Ye_PRL_2010,Kanomata_MaterSciForum_2008,Brown_JPCM_2006}. In these systems the Mn2 are antiferromagnetically coupled to Ni and Mn1 \cite{Pal_AIPProced_2014,Ye_PRL_2010,Kanomata_MaterSciForum_2008,Brown_JPCM_2006}, whereas Mn1 is ferromagnetically coupled to Ni. The temperature dependent X-ray diffraction (XRD) measurements are performed using synchrotron radiation of energy 18 KeV from room temperature to 10 K at Indian Beam line, Photon factory, KEK, Japan. 

The magnetic exchange parameter between first nearest neighbour Mn1-Mn2 of 16 atom unit cell of $x$ = 0.50 is calculated using experimentally obtained lattice parameters at different temperatures of $x$ = 0.44 on the basis of the idea of ising model where the energy of any spin system can be described by $E = - \sum\limits_{i \neq j} J_{ij}S_{i}S_{j}$. The theoretically adopted composition $x$ = 0.50, close to $x$ = 0.44, shows martensitic transformation in 16 atom unit cell structure. The nature of transformation in $x$ = 0.50 is similar to $x$ = 0.44. So, the behaviour of exchange parameter of $x$ = 0.50 can be attributed to the nature of magnetic exchange in $x$ = 0.44. The \textit{ab initio} calculations are performed using the PAW method as implemented in the VASP \cite{Kresse_PRB_2012} code within GGA for the exchange correlation functional. Monkhorst-Pack k-points mesh of 10 $\times$ 10 $\times$ 10 was used for calculation.

\section{Results and Discussion}
\begin{figure}[b]
\centering
\includegraphics[angle=0,width=9cm,height=12cm]{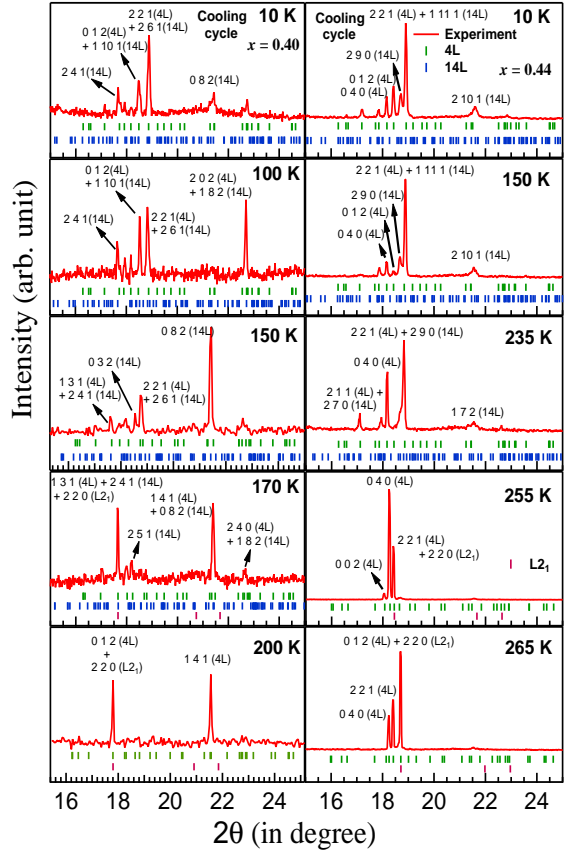}
\caption{Temperature variation of X-ray diffraction indexed by co-existing 4-layered and 14-layered orthorhombic crystal structure of $x$ = 0.40 and 0.44.}
\label{fig2}
\end{figure}
The temperature dependent crystal structure with fitting for $x$ = 0.40 and 0.44 are shown in fig. \ref{fig2}. The LeBail fitting is employed to extract out the contributing phase information and corresponding lattice parameters. It is noteworthy that MP has two co-existing orthorhombic structures, namely, 4-layered (4L) and 14-layered (14L) with space group $Pmma$. The AP has L2$_{1}$ cubic structure \cite{Sandeep_APL_2011}.  
\begin{table}[htb]
\centering
\caption{Lattice parameters and Volume of smallest unit cell of different structural phases after Bain transformation at various temperatures for $x$ = 0.40 and 0.44 compositions.}
\footnotesize
\resizebox{8cm}{!}{
\begin{tabular}{cccccc}
\hline
\hline
Compo- & Tempe- & Struc- & Lattice & Mn1-Mn2\\ 
sition& -rature& -tural  &parameter (\AA) & bond length (\AA)\\
($x$)& (K)& phase &  & \\
\hline
\multirow{7}{*}{0.40}& 300 & L2$_{1}$ & $a$  $\approx$ 5.99 & 3.00\\
&200 & 4L & $a$  $\approx$ 5.66, $b$  $\approx$ 8.68, $c$  $\approx$ 4.37& 2.92\\
&170 & 4L & $a$  $\approx$ 5.77, $b$  $\approx$ 8.71, $c$  $\approx$ 4.30& 2.88\\
& & 14L & $a$  $\approx$ 5.57, $b$  $\approx$ 27.49, $c$  $\approx$ 4.34 & 2.79\\
&100 & 4L & $a$  $\approx$ 5.64, $b$  $\approx$ 8.63, $c$  $\approx$ 4.38& 2.82\\
& & 14L & $a$  $\approx$ 5.60, $b$  $\approx$ 26.95, $c$  $\approx$ 4.32 & 2.80\\
&10 & 4L & $a$  $\approx$ 5.65, $b$  $\approx$ 8.64, $c$  $\approx$ 4.38& 2.82\\
& & 14L & $a$  $\approx$ 5.59, $b$  $\approx$ 26.99, $c$  $\approx$ 4.37 & 2.79\\
\hline
\multirow{7}{*}{0.44}& 300 & L2$_{1}$ & $a$  $\approx$ 5.98& 2.99\\
&265 & 4L & $a$ $\sim$ 5.95, $b$ $\sim$ 8.60, $c$ $\sim$ 4.32& 2.98\\
&245 & 4L & $a$  $\approx$ 5.65, $b$  $\approx$ 8.65, $c$  $\approx$ 4.40& 2.82\\
& & 14L & $a$  $\approx$ 5.59, $b$  $\approx$ 28.15, $c$  $\approx$ 4.40 & 2.79\\
&150 & 4L & $a$  $\approx$ 5.65, $b$  $\approx$ 8.64, $c$  $\approx$ 4.39& 2.82\\
& & 14L & $a$  $\approx$ 5.59, $b$  $\approx$ 28.59, $c$  $\approx$ 4.40 & 2.80\\
&10 & 4L & $a$  $\approx$ 5.61, $b$  $\approx$ 8.64, $c$  $\approx$ 4.39& 2.81\\
& & 14L & $a$  $\approx$ 5.58, $b$  $\approx$ 28.55, $c$  $\approx$ 4.41 & 2.79\\
\hline
\hline
\end{tabular}}
\label{Table2}
\end{table}            
The mixture of two structures in MP is, also, found in other Heusler alloys \cite{Kanomata_APL_2006,Khovaylo_PRB_2009,Ito_MMTA_2007}. The transformation from L2$_{1}$ cubic to 4L and 14L takes place by contraction along $c$-axis and elongation along $b$-axis according to Bain transformation. relation: $a_{ortho} \approx a_{cubic}, b_{ortho} \approx \frac{2}{\sqrt{2}} b_{cubic}, c_{ortho} \approx \frac{1}{\sqrt{2}} c_{cubic}$ and $a_{ortho} \approx a_{cubic}, b_{ortho} \approx \frac{7}{\sqrt{2}} b_{cubic}, c_{ortho} \approx \frac{1}{\sqrt{2}} c_{cubic}$, respectively. The lattice parameters for all the compositions and temperatures are summarized in Table \ref{Table2}. The contraction also along a-axis is observed in MP. The contraction in a-axis of 14L structure might be because transformation happens through 4L structure. The L2$_{1}$ structure initially transforms to 4L and then part of 4L transforms to 14L. The a-axis contraction of 4L structure is observed, mainly, when 4L and 14L structure co-exist. The intermartensitic transition is reported earlier for Ni-Mn-Ga, Ni-Mn-Fe-Ga, Ni-Mn-In and Ni-Fe-Ga \cite{Segui_AMR_2008,Koho_MSEA_2004,Huang_ActaMater_2013,Wang_PRB_2002,Pons_ActaMater_2007,Hamilton_ActaMater_2007}. The stability of the MP is achieved by transition to either 14M or non-modulated structure mainly through 10M structure. For present Ni-Mn-Sn alloys the martensitic transition to 14L structure is happening through 4L structure.  For Ni$_{2}$Mn$_{1.44}$Sn$_{0.56}$ the transition from L2$_{1}$ to 4L is reported earlier \cite{Brown_JPCM_2006}. The broad line-shape of exothermic and endothermic peaks (fig. \ref{fig1}) also imply existence of intermartensitic transition \cite{Krenke_PRB_2006}.

Evaluating the lattice parameters carefully, the percentage of phase fractions at various temperatures are obtained based on the ratio of integrated peak areas of individual 4L and 14L calculated intensity pattern and total calculated pattern.
\begin{figure}[htb]
\centering
\includegraphics[angle=0,width=7cm,height=11cm]{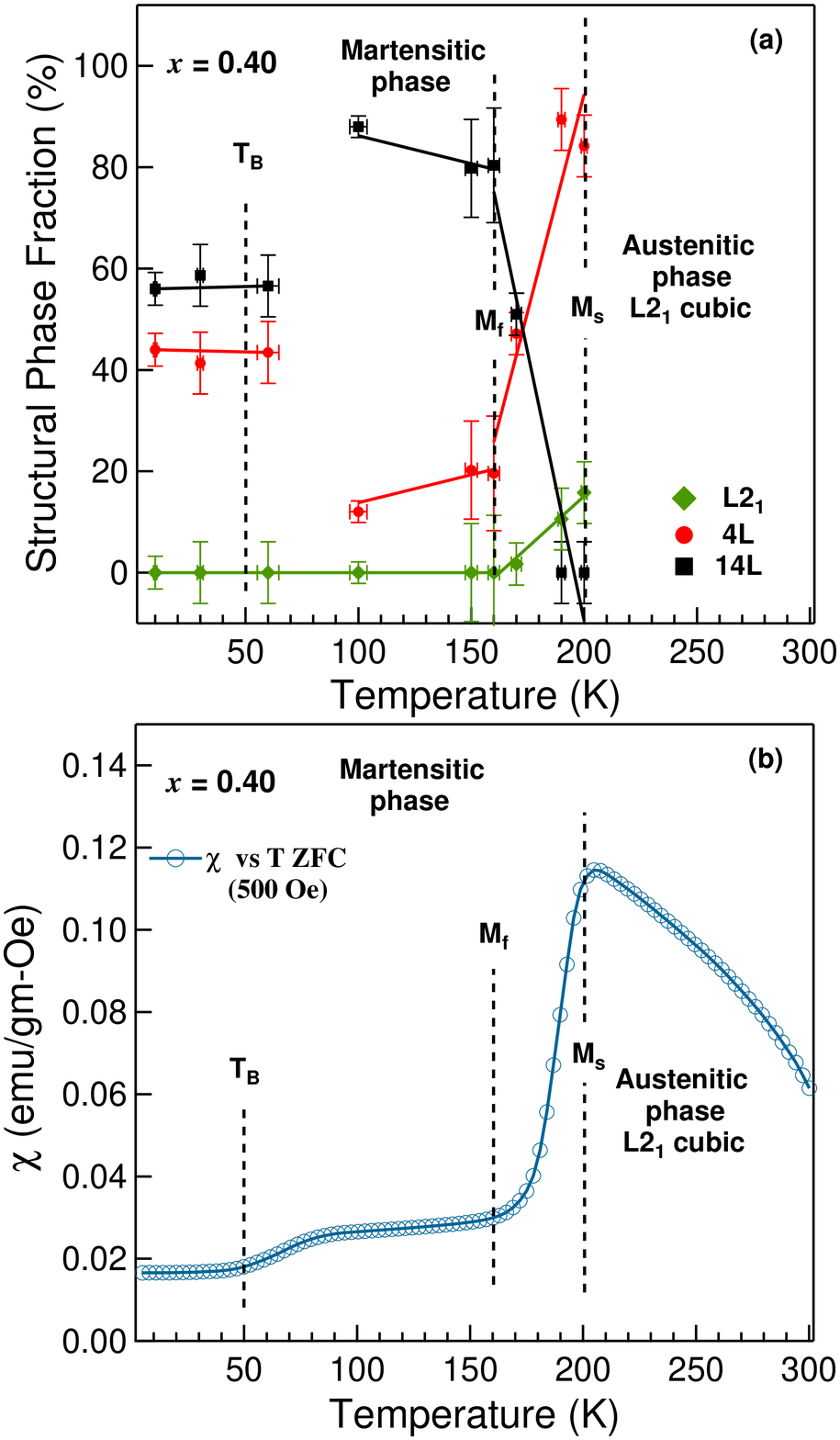}
\caption{(a) 4-layered and 14-layered structural phase fraction variation and (b) zero-field-cooled thermo-susceptibility variation in martensitic transition region, martensitic phase, and below T$_{B}$ region of $x$  = 0.40.}
\label{fig3}
\end{figure}      
Interestingly, the phase fraction of co-existing 4L and 14L structure varies as a function of temperature. The fig. \ref{fig3} (a) shows the phase fraction percentage of L2$_{1}$, 4L and 14L as a function of temperature for $x$ = 0.40.
Upon martensitic transition, initially, cubic L2$_{1}$ structure transforms to 4L structure. In between martensitic start (M$_{s}$) and martensitic finish (M$_{f}$) temperature range the 14L structure evolves at the cost of 4L structure. Between M$_{f}$ and spin freezing temperature T$_{B}$ (referred as T$_{f}$ in Ref. 39) the phase fraction is almost constant. Noteworthy that with further lowering of the temperature, the 4L phase fraction increases and 14L phase fraction decreases. The phase fraction of 4L and 14L is around 50 \% at 10 K. 
\begin{figure}[htb]
\centering
\includegraphics[angle=0,width=8cm,height=11cm]{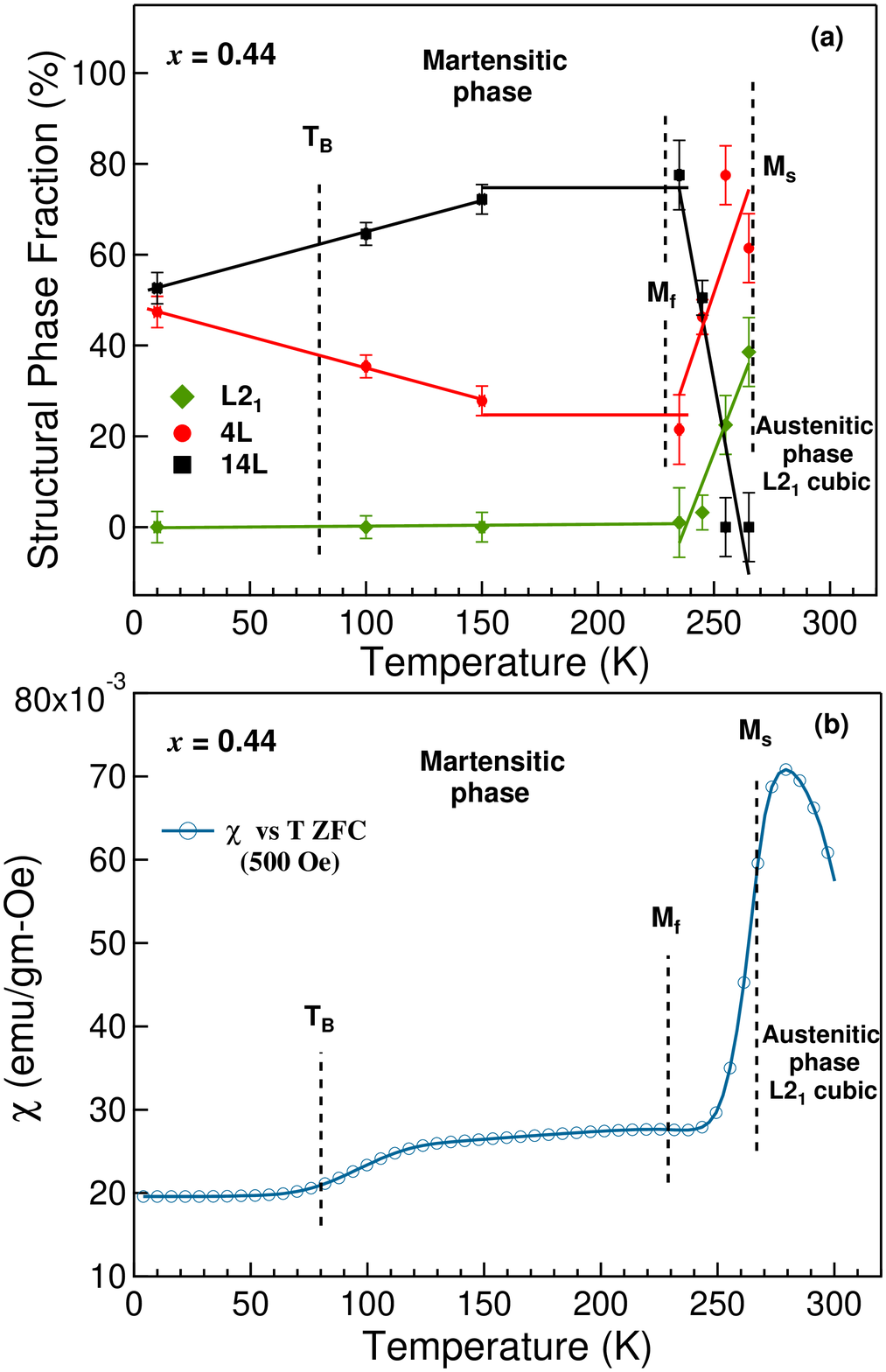}
\caption{(a) 4-layered and 14-layered structural phase fraction variation and (b) zero-field-cooled thermo-susceptibility variation in martensitic transition region, martensitic phase, and below T$_{B}$ region of $x$  = 0.44.}
\label{fig4}
\end{figure}

The formation energy per unit cell of 4L and 14L structures is calculated using experimental lattice constants of 14L and 4L structures at 150 K and 190 K, respectively, by \textit{ab initio} density functional theory. The formation energy per unit cell of 4L (-1.040315 eV) and 14L (-1.023121 eV) are very close to each other. Nevertheless, 4L requires less formation energy than 14L. Thus, initially, the martensitic transformation from parent phase to 4L structure occurs. To accommodate the stress accumulation by 4L structure, the transition to stacking sequence 14L occurs. This internal stress-related selectivity of intermartensitic transformation is also found in other systems like Ni-Mn-Ga, Ni-Mn-Fe-Ga, Ni-Mn-In, Ni-Fe-Ga etc. \cite{Segui_AMR_2008,Koho_MSEA_2004,Huang_ActaMater_2013,Wang_PRB_2002,Pons_ActaMater_2007,Hamilton_ActaMater_2007}. Since 4L is more favorable structure because it requires less formation energy than 14L, the 80\% phase fraction of 14L induces instability in the MP. Hence, to minimize the free energy of the system, the phase fraction of 4L structure increases once more. 

It is very important to note that the change in phase fraction of 4L and 14L occurs at the temperatures where magnetic phase change also occurs. The thermo-susceptibility behavior in zero field cooling (ZFC) for $x$ = 0.40 is shown in fig. \ref{fig3} (b). The lowering of susceptibility between M$_{s}$-M$_{f}$ indicates that the 14L structure possibly strengthens the AFM exchange interaction. The structural phase fraction is almost constant below M$_{f}$. Interestingly, below M$_{f}$ temperature the susceptibility, also, is almost constant. Around T$_{B}$ susceptibility drops again and 4L phase increases. So, the decrease in magnetization at T$_{B}$ may have structural correspondence. The changes of structural phase fractions of 4L and 14L and susceptibility at different temperatures for $x$ = 0.44, shown in fig. \ref{fig4} (a) and (b) respectively. They are very much similar to $x$ = 0.40. 

\begin{figure}[ht]
\centering
\includegraphics[angle=0,width=7.5cm,height=5cm]{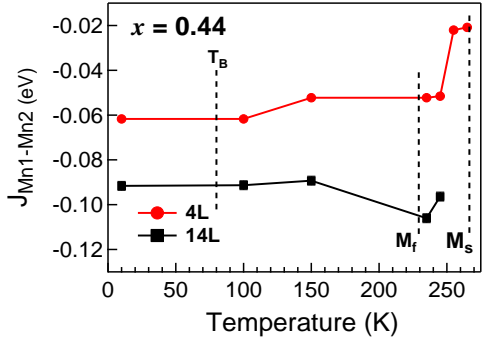}
\caption{Calculated magnetic exchange coupling constant (J) between first nearest neighbours Mn1-Mn2 of $x$ = 0.50 using experimentally obtained lattice parameters of $x$ = 0.44.}
\label{fig5}
\end{figure}

In Ni-Mn-Sn off-stoichiometric alloys the bond distance between Mn1 and Mn2 gives rise to the antiferromagnetism \cite{Krenke_Nature_2005,Pal_AIPProced_2014}. Thus, the bond distance between Mn1-Mn2 is deduced from the structural analysis as mentioned in Table \ref{Table2}. The exchange integral between Mn1-Mn2 (J$_{Mn1-Mn2}$) is calculated as a function of temperature (fig. \ref{fig5}). 
The calculation shows that AFM exchange interaction exists between Mn1-Mn2 in both 4L and 14L structures. Also, the AFM exchange interaction is stronger in 14L than 4L. This is also visible from the Mn1-Mn2 bond distances in Table \ref{Table2}. The 4L and 14L has bond distances less than 3 \AA ~ implying AFM interaction. Also, the Mn1-Mn2 bond distance of 14L is less than that of 4L. Thus, the AFM interaction in 14L is expected to be stronger than that in 4L. Thus, presence of two structural phases with different strength of AFM exchange interaction alongwith FM interaction gives rise to magnetically inhomogeneous phase. Moreover, it is known that the low temperature phase consists of various martensitic variants oriented in different directions derived from high symmetry cubic phase \cite{Zheng_ActaMater_2013}. The variants of orthorhombic 4L and 14L evolves randomly oriented. Thus, the average spin of the variants are also randomly oriented. With almost equal phase fraction of 4L and 14L, random orientation of their variants and spin, co-existence of FM and short-range AFM coupling, and different strength of AFM coupling might pin the FM moments by exchange bias to AFM spin moments. A diffuse AFM phase starts to develop below M$_{s}$ and becomes strong enough below T$_{B}$ (or T$_{f}$) to cause the spin freezing \cite{Giri_PRB_2009}. Thus, the re-entrant spin-glass like magnetic phase of Ni-Mn-Sn is obtained as reported in Ref. 39.

\section{Conclusion}

In the martensitic phase the co-existence of two crystal structures, change in crystal structure phase fraction with magnetic transition, co-existence of ferromagnetic and anti-ferromagnetic magnetic domains, and different strength of AFM coupling gives rise to structurally and magnetically frustrated martensitic phase that causes spin-glass-like behavior of Ni$_{2}$Mn$_{1+x}$Sn$_{1-x}$ ($x$ = 0.40 and 0.44) and lead to exchange bias phenomena under ZFC. Thus, the ZFC exchange bias in Ni-Mn-Sn alloys is related to the crystal structure of martensite.  

\section*{Acknowledgment}

We would like to thank DST-KEK for financial assistance to do experiment in Photon factory. We also wish to thank Dr. M. K. Mukhopadhyay and Mr. Satish Poddar for their help during experiment. We appreciate the support of Prof. Priya Mahadevan for theoretical calculation using VASP. Prof. B. N. Dev, Prof. G. P. Das and Prof. Shubham Majumdar are acknowledged for fruitful discussion. CM acknowledges DST, GoI for financial support vide reference no SR/WOS-A/PM-11/2016 under WOS-A.



\newpage

\end{document}